\documentclass[%
 aip,
 amsmath,amssymb,
 reprint,%
]{revtex4-1}

\usepackage{graphicx}
\usepackage{dcolumn}
\usepackage{bm}
\usepackage[ruled]{algorithm2e}

\usepackage[utf8]{inputenc}
\usepackage[T1]{fontenc}
\usepackage{mathptmx}
\usepackage{etoolbox}

\makeatletter
\def\@email#1#2{%
 \endgroup
 \patchcmd{\titleblock@produce}
  {\frontmatter@RRAPformat}
  {\frontmatter@RRAPformat{\produce@RRAP{*#1\href{mailto:#2}{#2}}}\frontmatter@RRAPformat}
  {}{}
}%
\makeatother
\begin{document}

\preprint{AIP/123-QED}

\title[ ]{Message passing approach to analyze the robustness of hypergraph}
\author{Hao Peng}

\author{Cheng Qian}%
\author{Dandan Zhao}%
\author{Ming Zhong}%
\author{Jianmin Han}%
\author{Runchao Li}%
\affiliation{ 
School of Computer Science and Technology, Zhejiang Normal University, Jinhua 321004, China}

\author{Wei Wang}
 \homepage{Email address of corresponding author:  wwzqbx@hotmail.com and wwzqbc@cqmu.edu.cn.}
\affiliation{%
School of Public Health, Chongqing Medical University, Chongqing, 400016, China
}

\date{\today}

\begin{abstract}
Hypergraph networks are closer to real life because they can reflect higher-order interactions, so researchers have begun using them to build models for real-world networks. The mean-field approach is the current tool for studying the percolation problem on hypergraph networks. However, we found that when there is a loop in the hypergraph network, the calculated results using this approach deviate from the real results. Therefore, in this paper, we rephrase the percolation on the hypergraph network as a message passing process, thus obtaining a message passing approach. Our proposed approach has been tested in several hypergraph networks with loops, and the experimental results are more accurate than those under the mean-field approach. This is helpful to analyze and understand the robustness of hypergraph networks with loops. In addition, we also specifically analyzed how four different types of loops affect the accuracy of the experiment. Our proposed message passing approach also provides another way to study percolation on hypergraph networks. 
\end{abstract}

\maketitle

\begin{quotation}
Since hypergraph networks can reflect the high-order interactions that exist in the real-world, the research on hypergraph theory has attracted more and more scientists' interest. In simple networks, scientists have found that as long as there are loops in the network (that is, not a local tree-like), the theoretical solution obtained by the mean-field approach always deviates from the true value. However, we found that this problem also exists on hypergraph networks with loops. Therefore, in this paper we propose a message passing approach, which solves the above problems well. In addition, we compared and analyzed the impact of four different types of loops on the accuracy of experimental results.
\end{quotation}

\section{\label{sec:level1}Inroduction}
As early as two decades ago, the problem of percolation on complex networks attracted the attention of a large number of scientists \cite{2010Networks,cohen2010complex,Dynamic,shao2009structure,newman2002spread,RevModPhys.74.47,nie2023pathogen,PhysRevE.66.016128}. Initially, the problem of percolation on a simple network was studied, and the dynamic process of cascading failure of the network was analyzed step by step. After simplifying the intermediate steps, the size of a giant connected cluster ($GCC$) can be found. With the development of time, in 2010, Buldyrev et al. \cite{buldyrev2010catastrophic} studied the percolation problem on the multilayer network with the help of the generating function. Since then, research papers on the multilayer network have continued to emerge \cite{huang2011robustness,dong2012percolation,cellai2013percolation,zhou2017security,xu2021breakdown}. The mean-field approach is generally used to study the percolation problem in a simple network. After randomly removing the $1-p$ proportion of nodes or edges in the network, the probability of a node belonging to the $GCC$ can be calculated. The average result of each experiment is taken as the final result. Mean-field approach averages the probability that all nodes belong to the $GCC$. The difference from the mean-field approach is that each node in the message passing approach will have a separate probability. Therefore, the mean-field approach is aimed at a network ensemble with a certain degree distribution, while the message passing approach is aimed at a specific network. 

The main idea of the message passing approach in simple networks is to establish an equation to solve the probability that a node belongs to the $GCC$ by introducing the probability that an edge belongs to the $GCC$. As another tool for analyzing network percolation, the advantage of message passing approach is that to obtain accurate results, the Monte-Carlo simulation algorithm must repeat the simulation in many experiments, which may take much time, and the final result still contains a statistical error. However, the message passing approach can obtain an exact solution only by one experiment on the tree-like network \cite{karrer2014percolation}. The fatal defect of the message passing approach is that its error is more obvious on the network with short loops, but this problem has been solved well by Newman et al. \cite{karrer2010message,cantwell2019message,newman2023message}.

In recent years, with the development of network science theory, network scientists propose to use hypergraph networks to characterize real-world network systems further \cite{coutinho2020covering,ghoshal2009random,pan2021predicting,lotito2022higher,bretto2013hypergraph,battiston2020networks}. Compared with simple graphs with pairwise interactions, hypergraphs have higher-order interactions and are better able to express some scenes that appear in daily life \cite{wang2022generalized,zhao2022higher,nie2022markovian,alvarez2022collective,zhao2022percolation,parastesh2022synchronization,majhi2022dynamics}. For example, when several co-authors jointly submit a paper, these authors have a cooperative relationship if the article is accepted. If the article is rejected, this relationship is no longer present. Therefore, we can see that the relationship of more than two nodes cannot be directly expressed by pairwise interaction, so the importance of hypergraph is self-evident.
\begin{figure*}[th]
\centering
\includegraphics[width=1.0\textwidth]{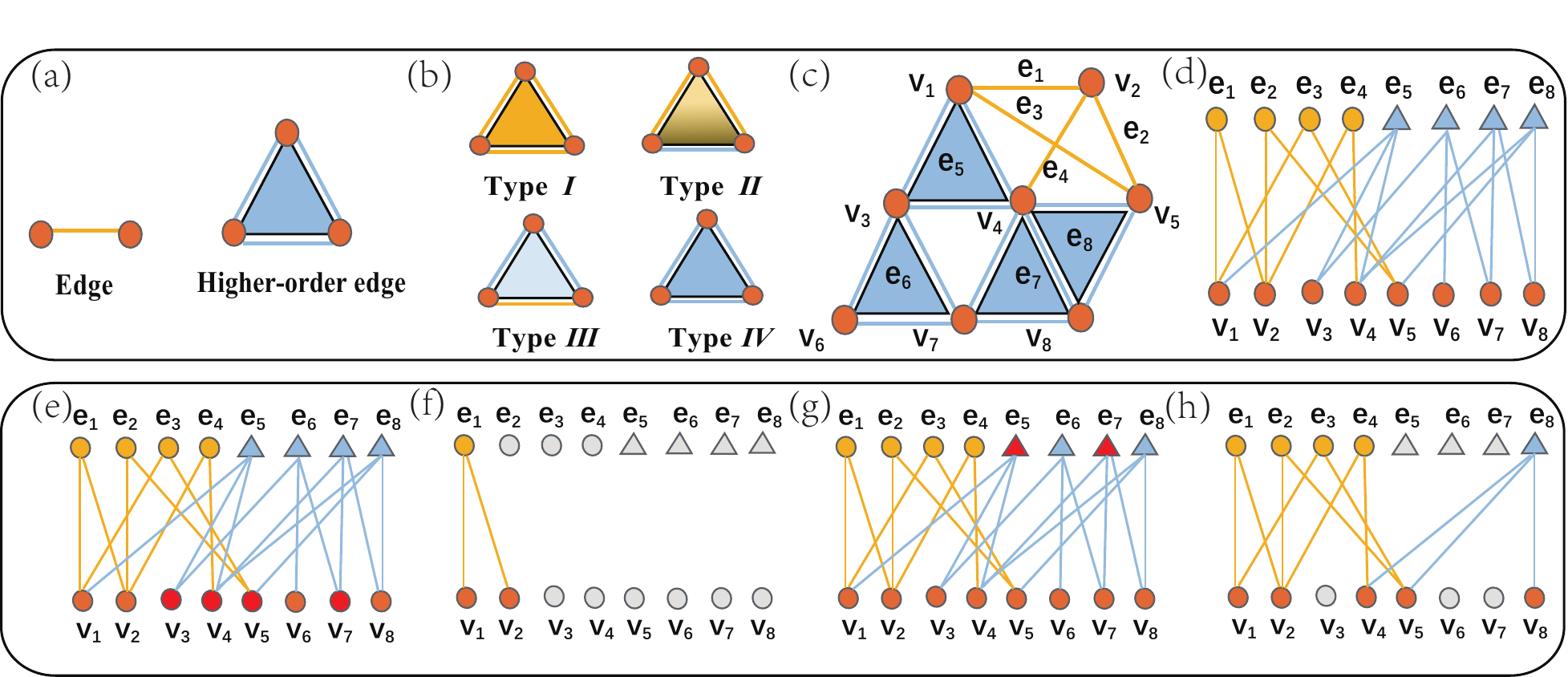}
\caption{Demonstrate the transformation of hypergraph into corresponding factor graph, site percolation and hyperedge percolation based on a factor graph. Panel (a) represents a schematic diagram of edges and higher-order edges, and edges are essentially equivalent to hyperedges with cardinality 2. Panel (b) shows four different types of loops formed by the borders between hyperedges, which are represented by Type $\uppercase\expandafter{\romannumeral1}$, Type $\uppercase\expandafter{\romannumeral2}$, Type $\uppercase\expandafter{\romannumeral3}$ and Type $\uppercase\expandafter{\romannumeral4}$ respectively. Panel (c) shows a hypergraph consisting of eight nodes and eight hyperedges, in which the number of loop types Type $\uppercase\expandafter{\romannumeral1}$, Type $\uppercase\expandafter{\romannumeral2}$, Type $\uppercase\expandafter{\romannumeral3}$ and Type $\uppercase\expandafter{\romannumeral4}$ are 1, 2, 1, and 1, respectively. Panel (d) is a schematic diagram of the factor graph corresponding to panel (c). It is represented as a bipartite graph to facilitate observing the percolation change in the factor graph. Panels (e)-(f) represent the site percolation process on the hypergraph. When $p^{[N]}$ = 0.5, node $v_{3}$, $v_{4}$, $v_{5}$, and $v_{7}$ are selected for removal, as indicated by the red markers. This will further cause node $v_{6}$ and $v_{8}$ ineffective. Currently, the final network size is 0.25 times the original network size. Panels (g)-(H) represent the hyperedge percolation process on the hypergraph. When $p^{[H]}$ = 0.25, hyperedge $e_{5}$ and $e_{7}$ are selected to be removed, as indicated by the red markers. This will further cause node $v_{3}$, $v_{6}$ and $v_{7}$ to leave the network, marked in grey. At this time, the final network size is 0.625 times of the original network size.}
\label{fig1}
\end{figure*}
Unlike a simple network comprising nodes and edges, a hypergraph network includes nodes and hyperedges. A random hypergraph network is the simplest one, characterized by the hyperdegree of the node, and the cardinality of hyperedge can obey any degree distribution. Therefore, any real-world network can use random hypergraph networks to build models, making studying random hypergraph networks more realistic. Aiming at the percolation problem on the hypergraph network, to facilitate theoretical analysis, Sun et al. \cite{sun2021higher} proposed to convert the hypergraph into a corresponding factor graph, regard a hyperedge as a factor node, and solve the site percolation or hyperedge percolation through a generating function with a self-consistent form. In the following paper, to distinguish it from our proposed message passing approach, we call this approach the mean-field approach. However, this approach's theoretical and Monte-Carlo simulation results are biased in hypergraph networks with loops. Therefore, we propose a message passing approach to hypergraph networks to study the robustness of hypergraph networks. Experiments on several hypergraph networks show that our approach is superior to the existing mean-field approach. Our contributions in this paper are as follows:

\begin{itemize}
\item We propose a message passing approach to study the percolation in hypergraph networks. The results obtained by using this approach in hypergraph networks with loops are more accurate than the existing mean-field approach. This also provides another perspective for studying robustness on hypergraph networks.
\item By constructing four different types of loops formed by the borders between hyperedges, we studied and analyzed the effect of different types of loops on the accuracy of the experiment in detail. The experimental results can help us design a more robust hypergraph network system.

\item We also studied how the giant connected cluster in the hypergraph network changes with the change of the proportion of nodes removed and the proportion of hyperedges removed in the hypergraph network, thus obtaining the relationship between the three.
\end{itemize}

\begin{figure*}[]
\centering
\includegraphics[width=0.8\textwidth]{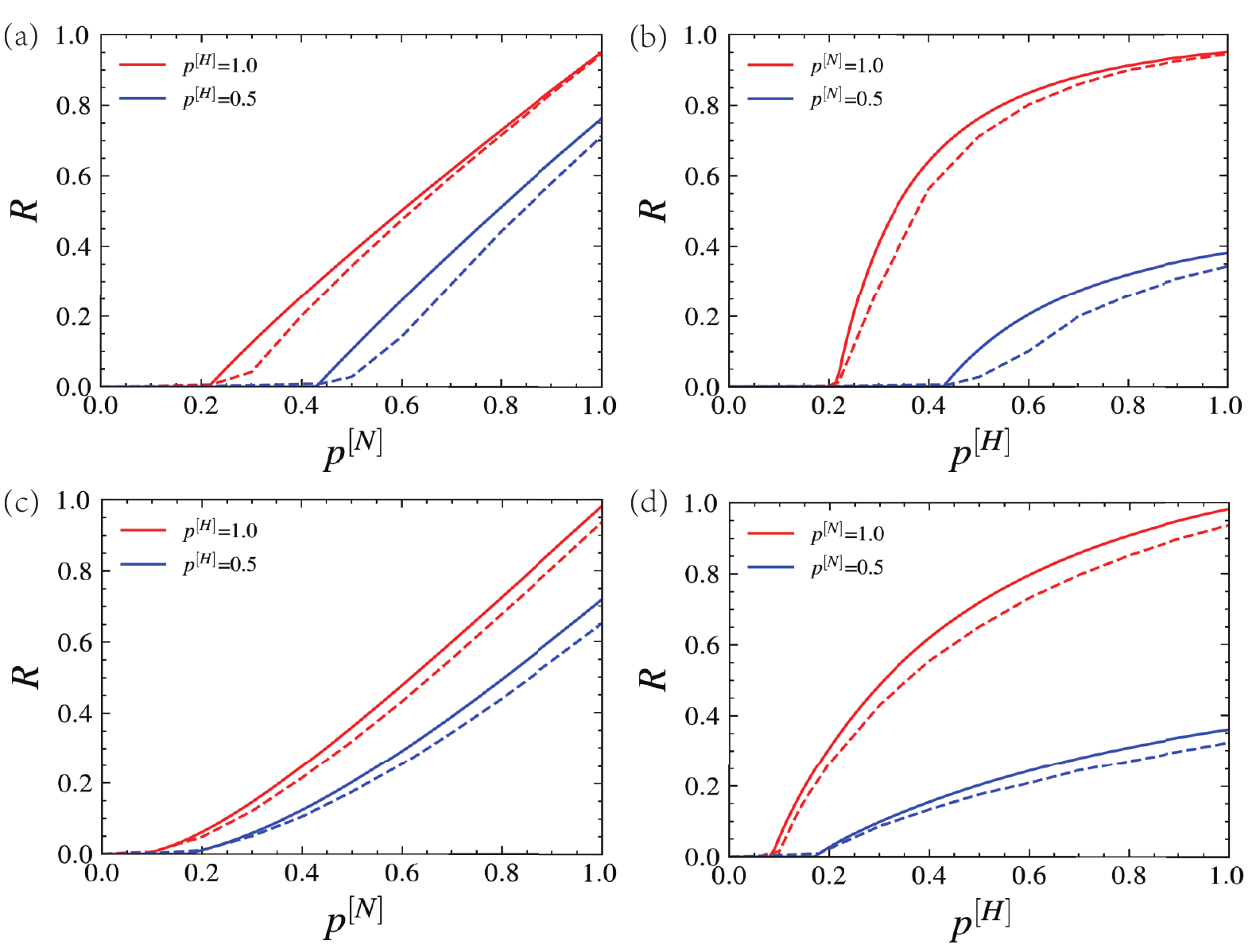}
\caption{Comparison of mean-field approach and Monte-Carlo Simulation in hypergraph network with Loop. Panels (a) and (b) respectively represent the site percolation and hyperedge percolation on the hypergraph network composed of 2,500 loops on the basis that the hyperdegree obeys Poisson distribution ( <k> = 3 ), and the hyperedge cardinality is always m = 2. Panels (c) and (d) respectively represent the site percolation and hyperedge percolation on the hypergraph network composed of 3,750 loops, where the hyperdegree obeys the power-law distribution ( <k> = 3 ) and the hyperedge cardinality is always m= 2. The solid line represents the results under the mean-field approach, and the dotted line represents the results under Monte-Carlo simulation. In Monte-Carlo simulation, N=5,000 and the results are averaged over 100 realizations. We can see that there are obvious differences between the two.}
\label{fig2}
\end{figure*}
\section{Model and Problem Preliminary}
In this paper, the model we use is the random hypergraph network model. In the hypergraph, the number of hyperedges that include a node is defined as the hyperdegree of the node, denoted by $k$; the number of nodes included in the hyperedge is defined as the cardinality of the hyperedge, denoted by $m$ \cite{peng2022disintegrate}. Nodes in the same hyperedge are entirely connected, so any hyperedge with cardinality greater than 2 has a loop inside. According to this property, hyperedges in hypergraphs can be divided into two types: hyperedges with a cardinality of two can be equivalent to edges (no loops), and hyperedges with cardinality greater than two are called high-order edges (with loops), as shown in Fig.\ref{fig1} (a). Therefore, except that the hyperedges with cardinality greater than two have loops inside, the borders between the hyperedges also form four different types of loops outside: a loop composed of three edges (Type $\uppercase\expandafter{\romannumeral1}$), a loop consisting of two edges and a higher-order edge (Type $\uppercase\expandafter{\romannumeral2}$), a loop composed of two higher-order edges and an edge (Type $\uppercase\expandafter{\romannumeral3}$), and a loop formed by three higher-order edges (Type $\uppercase\expandafter{\romannumeral4}$) , as shown in Fig.\ref{fig1} (b).

We first need to convert the hypergraph into a corresponding factor graph to facilitate the subsequent theoretical analysis. That is, a hyperedge is regarded as a factor node. If the hyperedge contains some nodes in the hypergraph, the factor nodes in the corresponding factor graph are connected to these nodes, as shown in Fig.\ref{fig1}(c) and (d). The details of the algorithm for building a random hypergraph network are as follows:

To make the hyperdegree of nodes and the cardinality of hyperedges obey the specified degree distribution in the hypergraph, this problem can be transformed into making the nodes and factor nodes follow the restricted degree distribution in the factor graph. The degree is determined in advance so that the total degree of the two types of nodes is equal, and they respectively obey the desired degree distribution. Finally, the unconnected stubs are randomly selected from the two types of nodes to connect until there are no remaining stubs in the two types of nodes. The constraint is that the same factor node is connected to the same node at most once.

The site percolation on the hypergraph refers to the study of the relationship between the $GCC$ and the node with occupancy probability $p^{[N]}$ after removing the $1-p^{[N]}$ ratio nodes in the network, as shown in Fig.\ref{fig1}(e)-(f), at this time $p^{[N]}$ = 0.5 because node $v_{3}$, $v_{4}$, $v_{5}$, and $v_{7}$ are removed. After cascading failures, the final network size accounts for 25 $\%$ of the original size. Hyperedge percolation refers to the study of the relationship between the $GCC$  and the hyperedge with occupancy probability $p^{[H]}$ after removing the $1-p^{[H]}$ proportion of hyperedges in the network, as shown in Fig.\ref{fig1}(g)-(h), where $p^{[H]}$ = 0.25, randomly select the hyperedge $e_{5}$ and $e_{7}$. After cascading failures, nodes $v_{3}$, $v_{6}$ and $v_{7}$ will be separated from the network, making the final network size account for 62.5 $\%$ of the original network.

The theoretical formula of the mean-field approach on the hypergraph is\cite{sun2021higher}:
\begin{equation}
\begin{array}{l}
\hat{S}=p^{[H]} \sum_{m} \frac{m}{\langle m\rangle} \hat{P}(m)\left[1-(1-S)^{m-1}\right], \\
S=p^{[N]} \sum_{k} \frac{k}{\langle k\rangle} P(k)\left[1-(1-\hat{S})^{k-1}\right] .
\end{array}
\label{1}
\end{equation}

Where $\hat{S}$ represents the probability of starting from a node and reaching a factor node belonging to the $GCC$ along an edge, and $S$ represents the probability of starting from a factor node and following an edge to a node belonging to the $GCC$. $<k>$ and $<m>$ represent the average hyperdegree and average hyperedge' cardinality, respectively. ${P}(k)$ and $\hat{P}(m)$ represent the degree distribution of nodes and factor nodes, respectively.

Then, the probability $R$ that a node belongs to the $GCC$ is as follows:
\begin{equation}
R=p^{[N]}\left[1-\sum_{k} P(k)(1-\hat{S})^{k}\right].
\label{2}
\end{equation}
 
Previous studies have shown that loops affect the accuracy of theoretically derived results in simple networks. However, in the real-world hypergraph network constructed in real life, it is only possible to partially guarantee that the corresponding factor graph is a tree-like structure; that is, there is no loop \cite{kim2022higher}. For example, in the friendship hypergraph network, the possibility of us becoming friends with friends of friends is often remarkably high. Therefore, we intentionally add loops to hypergraphs to explore whether the accuracy of the mean-field approach on hypergraphs is also affected by loops.

It should be noted that if there are hyperedges with cardinality greater than 3 in the simulation, it is very complicated to discuss the construction of the other three loops except for the loop of Type $\uppercase\expandafter{\romannumeral1}$. In order to better carry out quantitative analysis and simplify calculation, we all use hyperedges with cardinality size of 3 as higher-order edges during simulation. We first fix the cardinality of all hyperedges to 2 on the size of hypergraph network is $N=5,000$, and the hyperdegree obeyed the Poisson distribution, and the average hyperdegree on the network was 3. Theoretically, the number of hyperedges in the hypergraph network currently is 7,500 \cite{peng2022targeting}. We deliberately added 2,500 hyperedges with a cardinality of 2 so that each hyperedge could form a short loop (Type $\uppercase\expandafter{\romannumeral1}$). We used the mean-field approach and Monte-Carlo simulation to carry out site percolation and hyperedge percolation on the network, and the experimental results are shown in Fig.\ref{fig2}(a), (b).

Secondly, we are still in the hypergraph network size of 5,000, the cardinality size of all fixed hyperedges is all 2, the hyperdegree obeys the power-law distribution, the fixed average hyperdegree is 3, and $\lambda$ is 2.5 on the network. Similarly, the number of hyperedges in hypergraph network is 7,500 theoretically. On this basis, we deliberately add 3,750 hyperedges, so that each hyperedge can form a short loop (Type $\uppercase\expandafter{\romannumeral1}$). We also used the mean-field approach and Monte-Carlo simulation to carry out site percolation and hyperedge percolation on the network, and the experimental results are shown in Fig.\ref{fig2}(c), (d). For the construction algorithm of fixed average hyperdegree, refer to the algorithm of fixed average degree in Ref.\cite{gao2016effective}.

Through Fig.\ref{fig2}, it is easy to find that in hypergraph networks with loops, whether site percolation or hyperedge percolation, there is always a deviation between the analysis results obtained by the mean-field approach and the simulation. The reason for the deviation of the results under the mean-field approach is that the loop inside the hyperedge is ingeniously solved by converting a hypergraph into a factor graph, but the influence of the external loop on the experimental results still exists. To explore the effect of the loop type on the accuracy of the experiment, we set the number of nodes in the hypergraph network to $N$ = 10,000, the distribution of node hyperdegrees obeys the Poisson distribution, and the average hyperdegree is $<k>$ = 3, so the total number of hyperdegree is 30,000. Then we set the cardinality of hyperedges to 2 (edges), and the cardinality of hyperedges to 3 (higher-order edges) to 6,000 each, so the total cardinality of hyperedges is 30,000. In this way, we have built a hypergraph network model with both edges and higher-order edges. We mark the edges as 'label 1' and the higher-order edges as 'label 2'. If, in the process of building a hypergraph, the label between the two nodes is both 'label 1' and 'label 2', then we update the label to 'label 3'. Algorithm 1 shows its hypergraph construction process in detail.
\begin{algorithm}
\caption{Constructing hypergraphs with both edges and higher-order edges}
\KwIn{$N$ = 10,000, $\left \langle  \mathbf{k} \right \rangle $ = 3 and m = 2}
\KwOut{Hypergraphs with both edges and higher-order edges}
{
 $  \quad{Hypergraph \;H}$;\\
    $  \quad{A[N][N]=0}$;\quad//Initialize array.\\       
		$  \quad{N'\gets N*\left \langle k \right \rangle/ m}$;\quad//N' represents the total number of hyperedges.\\
		$  \quad m1=m2=N'/2$;\quad//m1 is the number of edges, m2 is the number of higher-order edges.\\
    \For{$i = 0 \to m1$}{ 
		$  \quad{a,b \gets random(0,N-1)}$; \quad// Select three nodes randomly from the network to build edges.\\ 
          $ \quad A[a][b]=A[b][a]=1;$ \quad//The number 1 assigned here is equivalent to the label 1 in the paper.\\ 
}
 \For{$i = m2 \to N'$}{ 
		$  \quad{a,b,c \gets random(0,N-1)}$; \quad//Select three nodes at random to build higher-order edges\\ 
          $ \quad A[a][b]=A[b][a]=A[a][c]=A[c][a]=A[b][c]=A[c][b]=2;$ \quad// The number 2 assigned here is equivalent to the label 2 in the paper. Please note that here you should first determine whether these array values are equal to 1. If so, assign the array value to 3 (label 3) to mark the shared edge.\\ 
}
\For{$j = 0 \to N$}{
       \quad \For{$k = 0 \to N$}{
	       \qquad \If{$A[j][k]!=0$}{	
           \qquad \quad  $ {H.add\_edge(j,k)}$;\\
           }
}
}  
 
    return  ${H}$;\\
}
\end{algorithm}

\begin{figure*}[thp]
\centering
\includegraphics[width=1.0\textwidth]{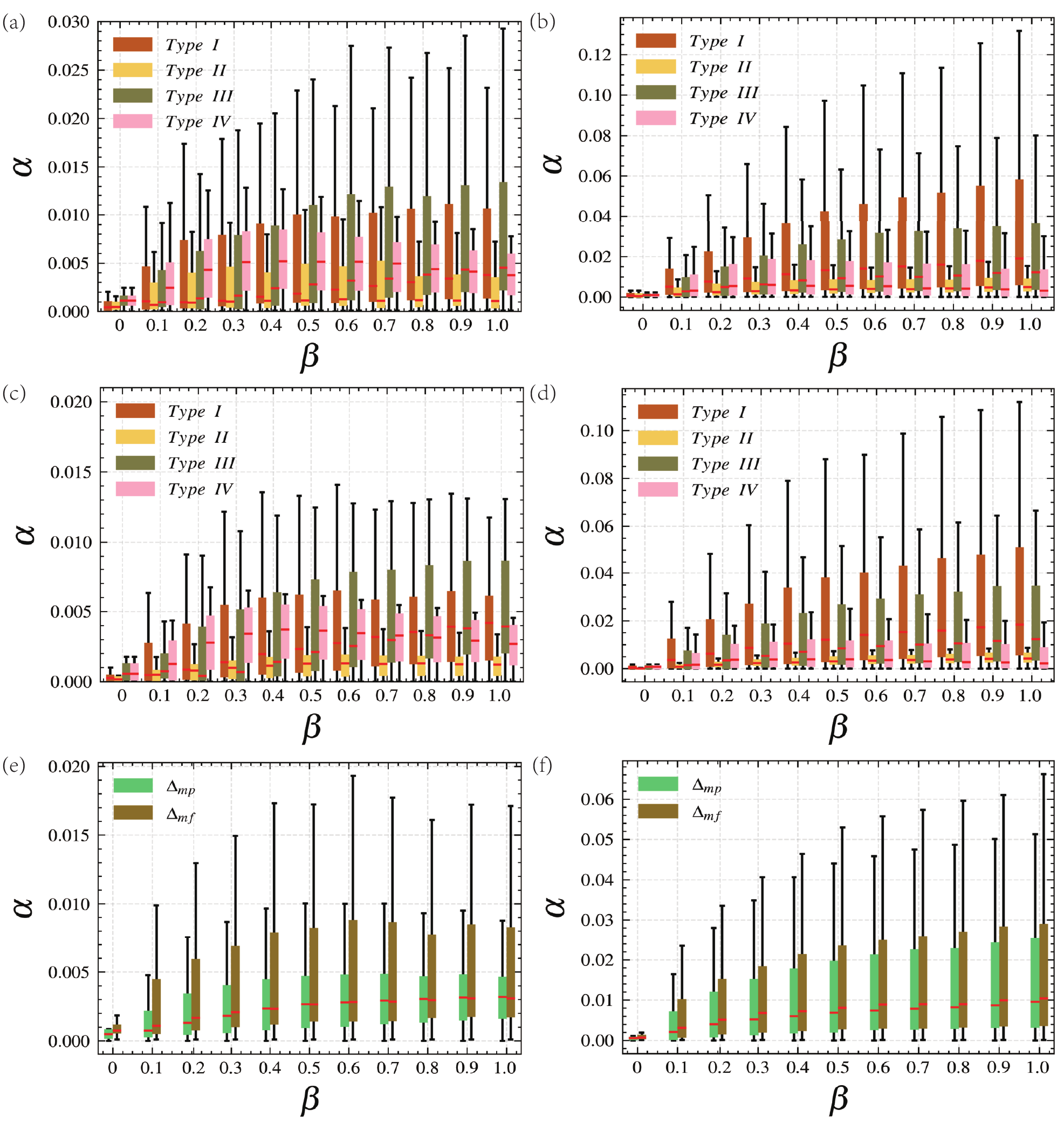}
\caption{The deviation comparison of the four types of loops under the mean-field approach and the message passing approach. The panels (a) and (b) respectively represent the variation of the deviation value after site percolation and hyperedge percolation, respectively calculated by the mean-field approach based on the hypergraph network composed of four different types of loops under the different proportion of added loops. The panels (c) and (d) respectively represent the variation of the deviation value after the occurrence of site percolation and hyperedge percolation through the message passing approach under the hypergraph network composed of four different types of loops with different added loop proportions. Under different ratios of added loops, the deviations generated by the four types of loop message passing approach and mean-field approach are averaged and recorded as $\Delta_{mp}$ and $\Delta_{mf}$, respectively. Panels (e) and (f) represent the changes of $\Delta_{mp}$ and $\Delta_{mf}$ after the occurrence of site percolation and hyperedge percolation, respectively. It can be seen from the experimental results that the message passing approach is superior to the mean-field approach in terms of both site percolation and hyperedge percolation. }
\label{fig3}
\end{figure*} 
 The purpose of labelling 'label 3' is to avoid selecting shared edges that are both edges and higher-order edges, thus eliminating the impact on the accuracy of experimental results. The original number of hyperedges in our hypergraph network is 12,000, and the ratio of the number of added edges forming a loop to the original number of hyperedges is set to $\beta$. Based on the hypergraph network model, we propose four strategies to construct four types of loops respectively. See algorithm 2 for details. We take the deviation between the result under the mean-field approach (or message passing approach) and the Monte-Carlo simulation result as the ordinate, which is represented by the symbol $\alpha$. Under different added edge proportions, the deviation calculation formula based on the mean-field approach is shown in Eq.(\ref{3}).
 \begin{algorithm}[b]
\caption{Build four different types of hypergraphs with loops}
\KwIn{$Hypergraph \; H, A[N][N], \beta*N'$(the number of loops) }
\KwOut{Four different types of hypergraphs with loops}
{ $ \ {i=0}$;\\
 \ \Switch{Loop type}{
   \  \quad    \uCase{Type $\uppercase\expandafter{\romannumeral1}$}{
       
 \  \quad \ \While{$(i<\beta*N')$}{
$\qquad \ {a,b,c \gets random(0,N-1)}$; \quad//Select three nodes randomly from the hypergraph network.\\ 
     \qquad \   \If{$(A[a][b]==1 \land A[b][c]==1 \land A[a][c]==0)$}{	
             $ \qquad \ \ {H.add\_edge(a,c)}$;  // Constitute a 'Type $\uppercase\expandafter{\romannumeral1}$' type of loop.     \\  
              $ \qquad \ \  A[a][c]=1$;\\
               $\qquad \ \  i++$;\\
           }
 }
}
     \  \quad    \uCase{Type $\uppercase\expandafter{\romannumeral2}$}{
       
 \  \quad \ \While{$(i<\beta*N')$}{
$\qquad \ {a,b,c \gets random(0,N-1)}$; \quad//Select three nodes randomly from the hypergraph network.\\ 
     \qquad \   \If{$(A[a][b]==1 \land A[b][c]==2 \land A[a][c]==0)$}{	
             $ \qquad \ \ {H.add\_edge(a,c)}$; // Constitute a 'Type $\uppercase\expandafter{\romannumeral1}$' type of loop.     \\  
              $ \qquad \ \  A[a][c]=1$;\\
               $\qquad \ \  i++$;\\
           }
 }
}

\  \quad    \uCase{Type $\uppercase\expandafter{\romannumeral3}$}{     
 \  \quad \ \While{$(i<\beta*N')$}{
$\qquad \ {a,b,c \gets random(0,N-1)}$; \quad//Select three nodes randomly from the hypergraph network.\\ 
     \qquad \   \If{$(A[a][b]==2 \land A[b][c]==2 \land A[a][c]==0)$}{	
             $ \qquad \ \ {H.add\_edge(a,c)}$; // Constitute a 'Type $\uppercase\expandafter{\romannumeral1}$' type of loop.     \\  
              $ \qquad \ \  A[a][c]=1$;\\
               $\qquad \ \  i++$;\\
        }
  }
}
 \  \quad  \uCase{Type $\uppercase\expandafter{\romannumeral4}$}{

\  \quad  \ \While{$(i<\beta*N')$}{
      $ \qquad \ {a,b,c,d \gets random(0,N-1)}$; \quad//Select four nodes randomly from the hypergraph network.\\ 
     \qquad \    \If{$(A[a][b]==2 \land A[b][c]==2 \land A[a][c]==0 \land A[a][d]!=1 \land A[c][d]!=1)$}{	
				 \qquad \ \	A[a][c]=A[c][a]=A[a][d]=A[d][a]=A[c][d]=A[d][c]=2;  \quad// Build a high-order edge.    \\ 
             $  \qquad \ \ {H.add\_edge(a,c)}$;  // Constitute a 'Type $\uppercase\expandafter{\romannumeral4}$' type of loop.     \\ 
    $\qquad \ \  i++$;\\ 
           }

}}
 }
    return  ${H}$;\\
}
\end{algorithm}

\begin{equation}
\alpha=\int_{0}^{1}\left|S_{\mathrm{sm}}-S_{\mathrm{mf}}\right| dp
\label{3}
\end{equation}

where $S_{\mathrm{sm}}$ is the real result under simulation, and $S_{\mathrm{mf}}$ is the result obtained under the mean-field approach. $p$ is the proportion of retained nodes or hyperedges\cite{jones2022improving}.

Fig.\ref{fig3} (a), (b) show the variation of the deviation value of the four types of loops formed by the above four algorithms under different added edge proportions under site percolation and hyperedge percolation, respectively. We can see from the median in the box chart that, in site percolation, 'Type $\uppercase\expandafter{\romannumeral4}$' type of loop has the greatest impact on the accuracy of the network, 'Type $\uppercase\expandafter{\romannumeral3}$' type of loop has the second most impact, and 'Type $\uppercase\expandafter{\romannumeral2}$' type of loop has the least impact. In hyperedge percolation, 'Type $\uppercase\expandafter{\romannumeral1}$' type of loop has the greatest impact on the accuracy of the network, 'Type $\uppercase\expandafter{\romannumeral3}$' type of loop has the second most impact, and 'Type $\uppercase\expandafter{\romannumeral2}$' type of loop has the least impact.

\section{Analytical Solution}
	\label{C}
Based on the above existing problems, the existing approach cannot accurately solve hypergraph networks with loops, so the main contribution of this paper is to propose a message passing approach to hypergraph networks to solve such problems effectively. In this section, we detail our proposed workaround. Consider two types of nodes in the factor graph: nodes and factor nodes. Therefore, there are two cases when considering the connection of edges: the node's connection to the factor node and the connection of the factor node to the node.
\begin{figure*}[t]
\centering
\includegraphics[width=1.0\textwidth]{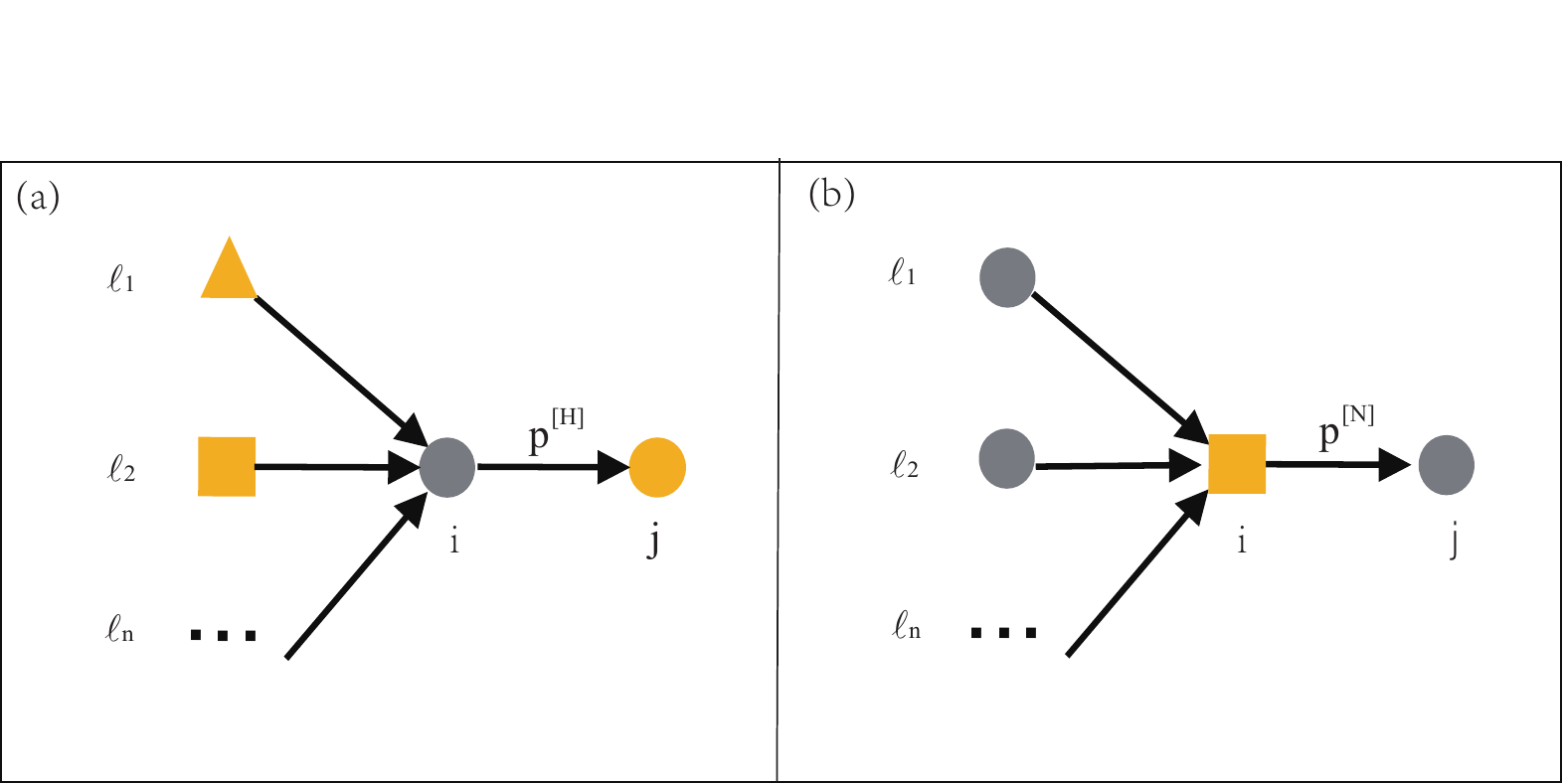}
\caption{Schematic diagram of message passing on hypergraph network. Grey represents nodes, and orange represents factor nodes. Panel (a) indicates that after the factor node is removed with a probability of 1 - $p^{[H]}$, the probability $\hat{\tau  } _{i \rightarrow j}$ of an edge of node $i$ reaching the factor node $j$ in the $GCC$ is related to the probability of other neighbour factor nodes of node $i$ to node $i$. Panel (b) indicates that after the node is removed with a probability of 1 - $p^{[N]}$, the probability $\tau_{i \rightarrow j}$ of an edge of factor node $i$ reaching the node $j$ in the $GCC$ is related to the probability of other neighbour nodes of factor node $i$ to factor node $i$.}
\label{fig4}
\end{figure*}

In essence, the "message" passed between the node and the factor node through the edge refers to the probability. We define the probability of random removal of nodes as $1-p^{[N]}$, and the probability of random removal of factor nodes (hyperedges) as $1-p^{[H]}$. 
 let the probability of reaching a factor node $j$ belonging to the $GCC$ along an edge of a node $i$ be $\hat{\tau } _{i\to j} $. Similarly, let the probability of reaching node $j$ belonging to the $GCC$ along an edge of a factor node $i$ be $\tau  _{ i\to j}$, as shown in Fig.\ref{fig4}. We can get the following self-consistent equation: 
\begin{equation}
\begin{aligned}
\hat{\tau  } _{i \rightarrow j}= p^{[H]}\left[1-\prod_{\ell \in \mathcal{N} (i) \backslash j}\left(1-\tau_{\ell \rightarrow i}\right)\right],\\
\tau_{i \rightarrow j}=p^{[N]}\left[1-\prod_{\ell \in \mathcal{N} (i) \backslash j}\left(1-\hat{\tau  }_{\ell \rightarrow i}\right)\right].
\end{aligned}
\label{4}
\end{equation}
where $\mathcal{N} (i) $ represents the neighborhood of node $i$.

At this time, the probability $R_{i}$ that node $i$ belongs to the $GCC$ is:

\begin{equation}
R_{i}=p^{[N]}\left[1-\prod_{\ell \in \mathcal{N} (i)}\left(1-\hat{\tau }_{\ell \rightarrow i}\right)\right].
\label{5}
\end{equation}
Therefore, the probability $R$ of a node belonging to the $GCC$ is:
\begin{equation}
R=\frac{p^{[N]}}{N} \sum_{i=1}^{N}\left[1-\prod_{\ell \in \mathcal{N} (i)}\left(1-\hat{\tau }_{\ell \rightarrow i}\right)\right].
\label{6}
\end{equation}

The critical threshold is obtained by linearizing the message passing Eq.(\ref{4}), obtaining
\begin{equation}
\begin{array}{l}
\hat{\tau }_{i \rightarrow j} =  p^{[H]}\sum_{\ell \in  \mathcal{N}(i)} \mathcal{M}_{\ell i \rightarrow i j} \tau_{\ell \rightarrow i},\\
\tau_{i \rightarrow j} = p^{[N]}\sum_{\ell \in  \mathcal{N}(i)} \mathcal{M}_{\ell i \rightarrow i j}\hat{\tau }_{\ell \rightarrow i}.
\end{array}{}  
\label{7}
\end{equation}

Here $\mathcal{M}$ contains elements
\begin{equation}
\mathcal{M}_{\ell i \rightarrow i j}=a_{\ell i} a_{i j}\left(1-\delta_{\ell j}\right),
\label{8}
\end{equation}
where $a$ represents the adjacency matrix of the network, and $\delta$ represents the Kronecker delta function.

In this way, solving this system of linear Eq.(\ref{7}), we get
\begin{equation}
\tau_{\ell \rightarrow i} =p^{[N]}\sum_{\ell^{\prime} \in N(\ell)} \mathcal{M}_{\ell^{\prime} \ell \rightarrow \ell i}   \hat{\tau }_{\ell^{\prime} \rightarrow \ell}.
\label{9}
\end{equation}

When the largest eigenvalue $\Lambda(\mathcal{E})$ of the modified non-backtracking matrix $\mathcal{E}$ is equal to 1 \cite{bianconi2021message}, the process has a threshold, that is
\begin{equation}
\Lambda(\mathcal{E})=1.
\label{10}
\end{equation}

Therefore, by substituting Eq.(\ref{9}) into Eq.(\ref{7}), we can obtain the modified non-backtracking matrix $\mathcal{E}$ from the non-backtracking matrix $\mathcal{M}$, so that we can further obtain the critical point, where $\mathcal{E}$ is given by
 
	\begin{equation}
\mathcal{E}_{\ell^{\prime} \ell \rightarrow i j}=p^{[N]}p^{[H]}\mathcal{M}_{\ell^{\prime} \ell \rightarrow \ell i}\mathcal{M}_{\ell i \rightarrow i j}. 
\label{11}
\end{equation}

\section{Results}
In this section, we first verify the accuracy of the message passing approach on the hypergraph without loops, and then we also verify that the approach is still effective on the hypergraph network with loops. In addition, we also studied the influence of network size and loop type on the accuracy of experimental results in the context of hypergraph networks with loops. Finally, we verify our approach on three real-world hypergraph networks. To verify the accuracy of our approach, we will repeat the Monte-Carlo simulation experiment 100 times each time, and the final result is the average of these 100 times. Take the final result as the true value. All experiments were carried out on a personal computer with 32G memory and 3.70GHz AMD Ryzen 9 5900X 12-Core Processor.

\begin{figure*}[t]
\centering
\includegraphics[width=1.0\textwidth]{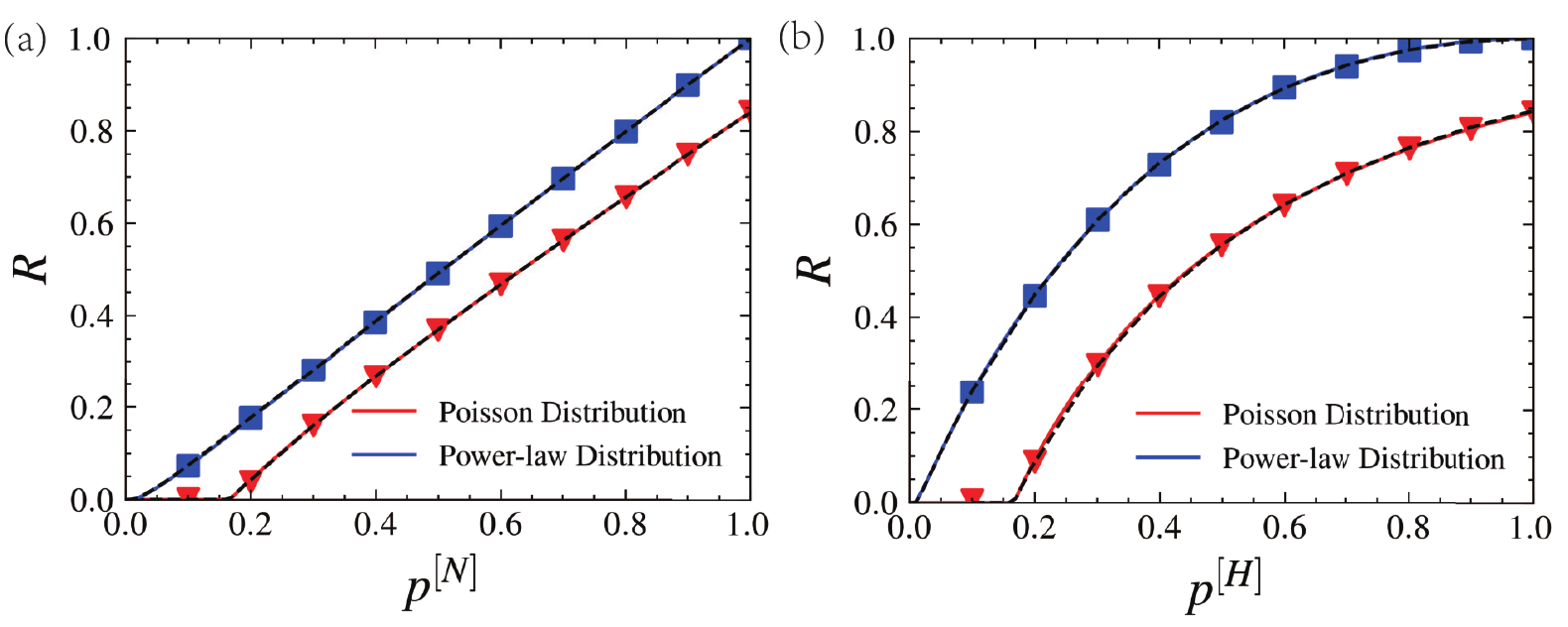}
\caption{Message passing on a hypergraph network without loops. Panel (a) and panel (b) are, respectively, site percolation and hyperedge percolation on random hypergraph networks and scale-free hypergraph networks. The solid line represents the result under the mean-field approach. The dotted line represents the result under the message passing approach, and the symbol represents the result under simulation. It can be seen that the three fit well.}
\label{fig5}
\end{figure*}
\subsection{Loopless network}

Here, we consider applying our message passing approach to a random hypergraph network, where the hyperdegree of nodes and the cardinality of hyperedges can obey the specified degree distribution. When the average hyperdegree and the average hyperedge cardinality are small, and the network size is large, the possibility of loops in the network is close to zero. This is used to verify the accuracy of the message passing approach on the loopless hypergraph network. First, consider that the hyperdegree of nodes and the cardinality of hyperedges obey the Poisson degree distribution, where the Poisson degree distribution is:

\begin{equation}
\begin{array}{l}
p(k)=\frac{e^{-\langle k\rangle}\langle k\rangle^{k}}{k!},\\
p(m)=\frac{e^{-\langle m\rangle}\langle m\rangle^{m}}{m!}.
\end{array}
\label{12}
\end{equation}
\begin{figure*}[t]
\centering
\includegraphics[width=1.0\textwidth]{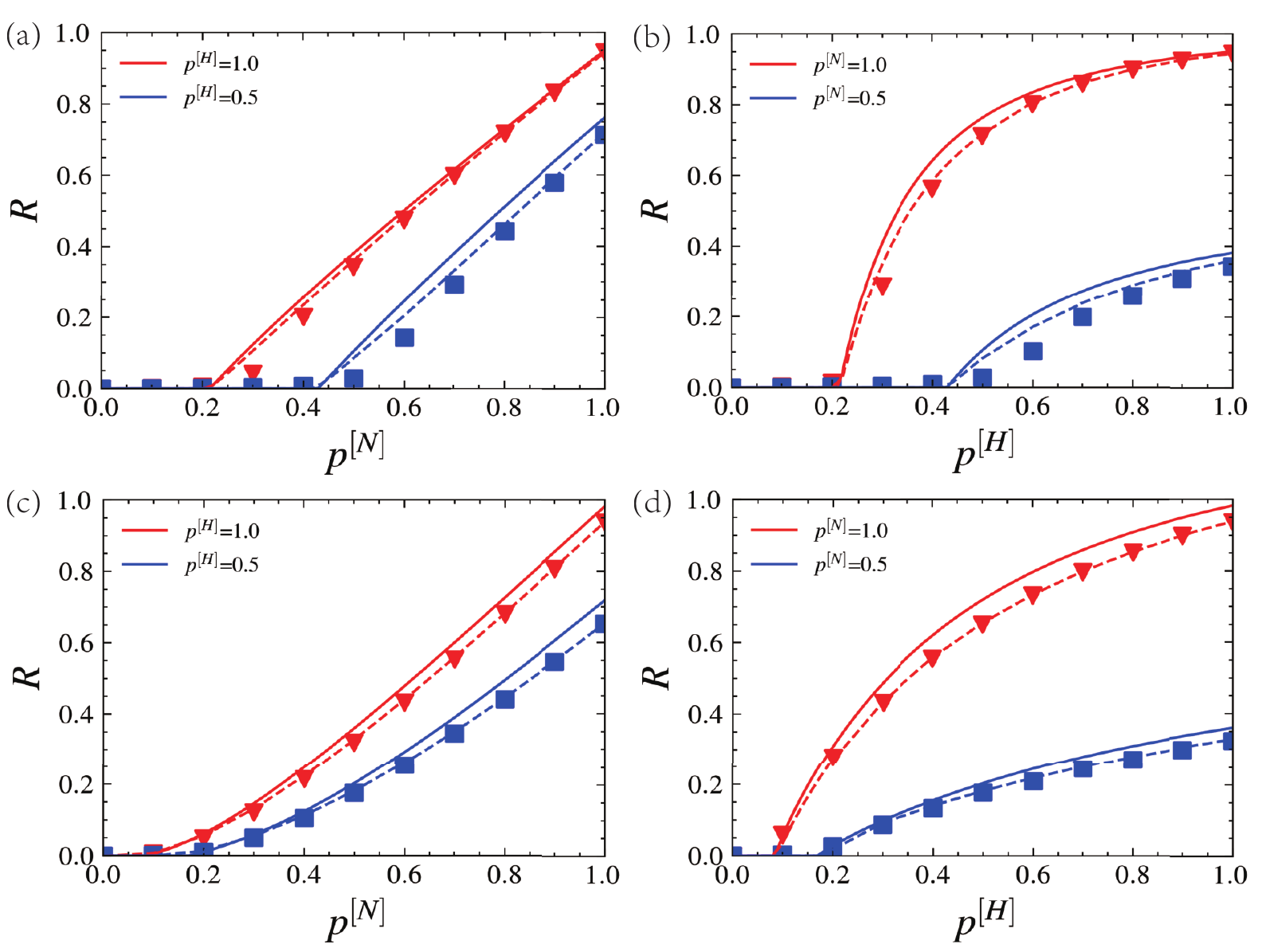}
\caption{Comparison of results of mean-field approach, message passing approach and Monte-Carlo simulation on hypergraph networks with Loops. Based on the hypergraph network model in Fig.2, we again use our proposed message passing approach to conduct experiments. The solid line represents the result under the mean-field approach. The dotted line represents the result under the message passing approach, and the symbol represents the result under simulation. The experimental results show that for the percolation problem that occurs on a hypergraph network with loops, the results obtained by using the mean-field approach deviate from the simulation results, but the results obtained by the message passing approach can correspond well with the simulation results. }
\label{fig6}
\end{figure*}

Where $<k>$ and $<m>$ are nodes' average hyperdegree size and hyperedges' average cardinality size, respectively, we experiment on a hypergraph network with 10,000 nodes, where $<k>$ = 2 and $<m>$ = 3. The experimental results of site percolation and hyperedge percolation on the random hypergraph network are shown in Fig.\ref{fig5}. The results show that the message passing approach is as good as the mean-field approach on the network without a loop.

Second, consider applying our message passing approach to a random hypergraph network in which both the hyperdegree of nodes and the cardinality of hyperedges obey a power-law degree distribution, where the power-law degree distribution is:

\begin{equation}
\begin{array}{l}
p(k)=\frac{(k+1)^{1-\lambda_k}-k^{1-\lambda_k}}{(D_k+1)^{1-\lambda_k}-d_k^{1-\lambda_k}},\\
p(m)=\frac{(m+1)^{1-\lambda_m}-m^{1-\lambda_m}}{(D_m+1)^{1-\lambda_m}-d_m^{1-\lambda_m}}.
\end{array}
\label{13}
\end{equation}
Among them, $D_k$, $d_k$, $D_m$, and $d_m$ are the maximum hyperdegree, minimum hyperdegree, and maximum cardinality and minimum cardinality of the hyperedge, respectively. We also conducted experiments on hypergraph networks with $N$ = 10,000 nodes, and for convenience, we set the number of hyperedges to the same number of nodes. Therefore, we take $D_k$ = $D_m$ = $\sqrt{ N}$, $d_k$ = $d_m$ =2 and $\lambda_k$ = $\lambda_m$ = $\lambda$ = 2.7. The experimental results of site percolation and hyperedge percolation on the scale-free hypergraph network are shown in Fig.\ref{fig5}. The results also show that the message passing approach is as good as the mean-field approach on the network without a loop.

\subsection{Loop network}

In this subsection, first of all, let's go back to the problem mentioned in Fig.\ref{fig2} that there is a deviation between the mean-field approach and the results under the simulation under the hypergraph network with loops. In this context, we use our proposed message passing approach to conduct experiments under the same hypergraph network, and the experimental results are shown in Fig.\ref{fig6}. It is easy to see that in the case of obvious deviation between the results of the mean-field approach and the results of Monte-Carlo simulation, the results of the message passing approach are consistent with the results of Monte-Carlo simulation. Therefore, our approach also has good accuracy on the hypergraph network with loops. In addition, we can also see that when the size of GCC is greater than 0, the robustness of the hypergraph network becomes stronger with the increase of the proportion of reserved nodes or hyperedges.

\begin{figure*}[t]
\centering
\includegraphics[width=1.0\textwidth]{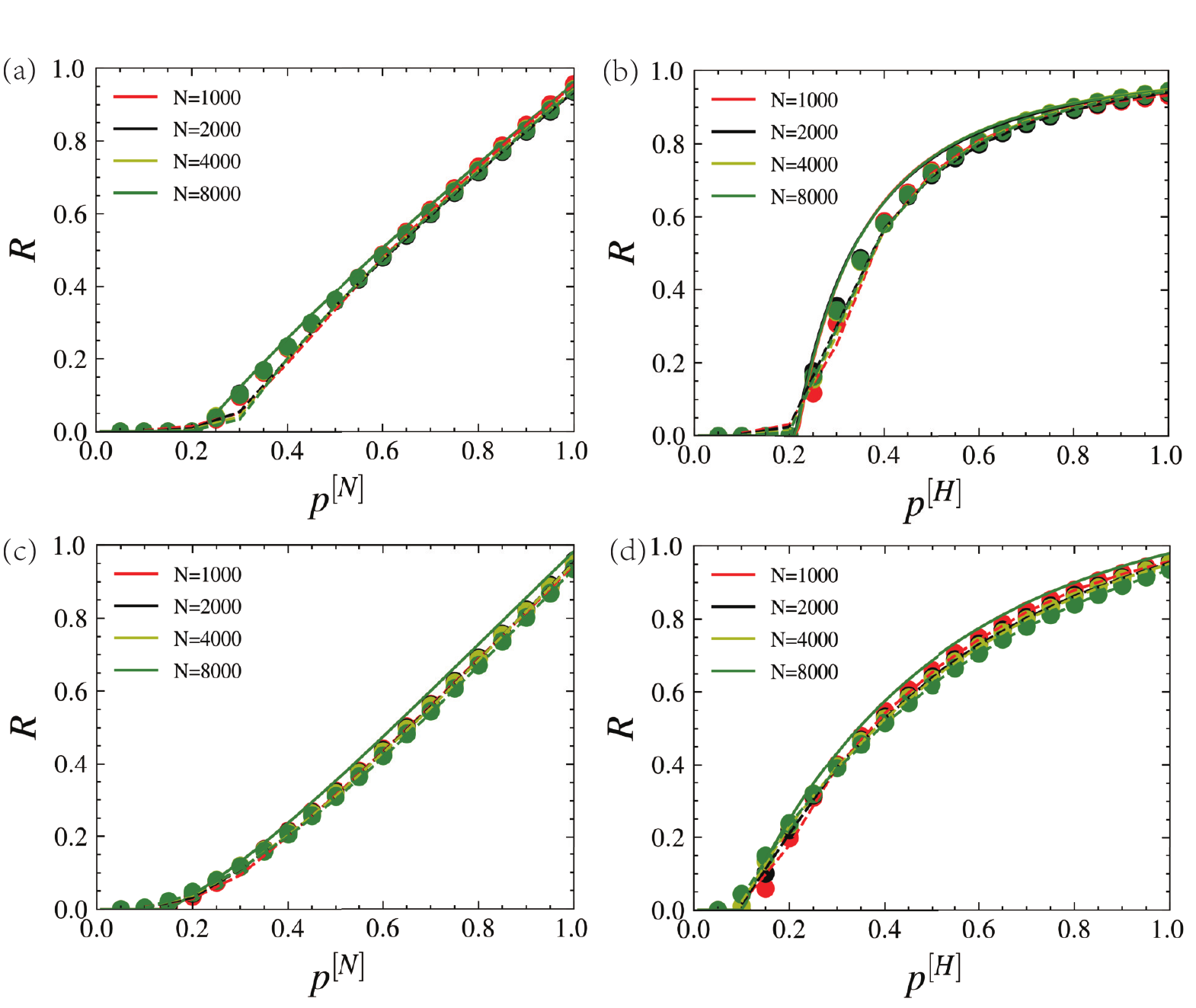}
\caption{Explore the influence of different network sizes on the robustness of hypergraph networks with loops. Panel (a) and panel (b) show that based on the network formed by hyperdegree obeying Poisson distribution ($<k>$ = 3) and hyperedge cardinality being identical to 2, the site percolation and hyperedge percolation on the hypergraph network formed by adding hyperedges accounting for 1/3 of the total number of hyperedges are added. Panel (c) and panel (d) show that based on the network formed by $\lambda$ = 2.5, fixed average hyperdegree ($<k>$ = 3) and hyperedge cardinality being identical to 2, the site percolation and hyperedge percolation on the hypergraph network formed by adding hyperedges accounting for 1/2 of the total number of hyperedges are added. The solid line represents the theoretical solution based on the mean-field, the dotted line represents the theoretical solution based on message passing, and the scattered point represents the simulation solution. Different colours represent different network scales.}
\label{fig7}
\end{figure*}

Secondly, we study the influence of the size of hypergraph network on the accuracy of message passing approach. We carried out percolation experiments on hypergraph networks with multiple loops with different underlying topologies. Take the network size as an independent variable, and control other variables to remain unchanged, as shown in Fig.\ref{fig7}. Through the coincidence of solid lines representing the results of the mean-field approach under different network sizes, we can conclude that the results obtained by the mean-field approach are very stable. In addition, we can also see that no matter how the size of the hypergraph network changes, there will be deviations in the application of the mean-field approach to the hypergraph network with loops. However, the accuracy of the message passing approach is still very good regardless of how the network size changes.

Next, we quantitatively analyze the deviation between our proposed approach and the Monte-Carlo simulation through Eq.(\ref{14}).

\begin{equation}
\alpha=\int_{0}^{1}\left|S_{\mathrm{sm}}-S_{\mathrm{mp}}\right| dp
\label{14}
\end{equation}

\begin{figure*}[t]
\centering
\includegraphics[width=1.0\textwidth]{ 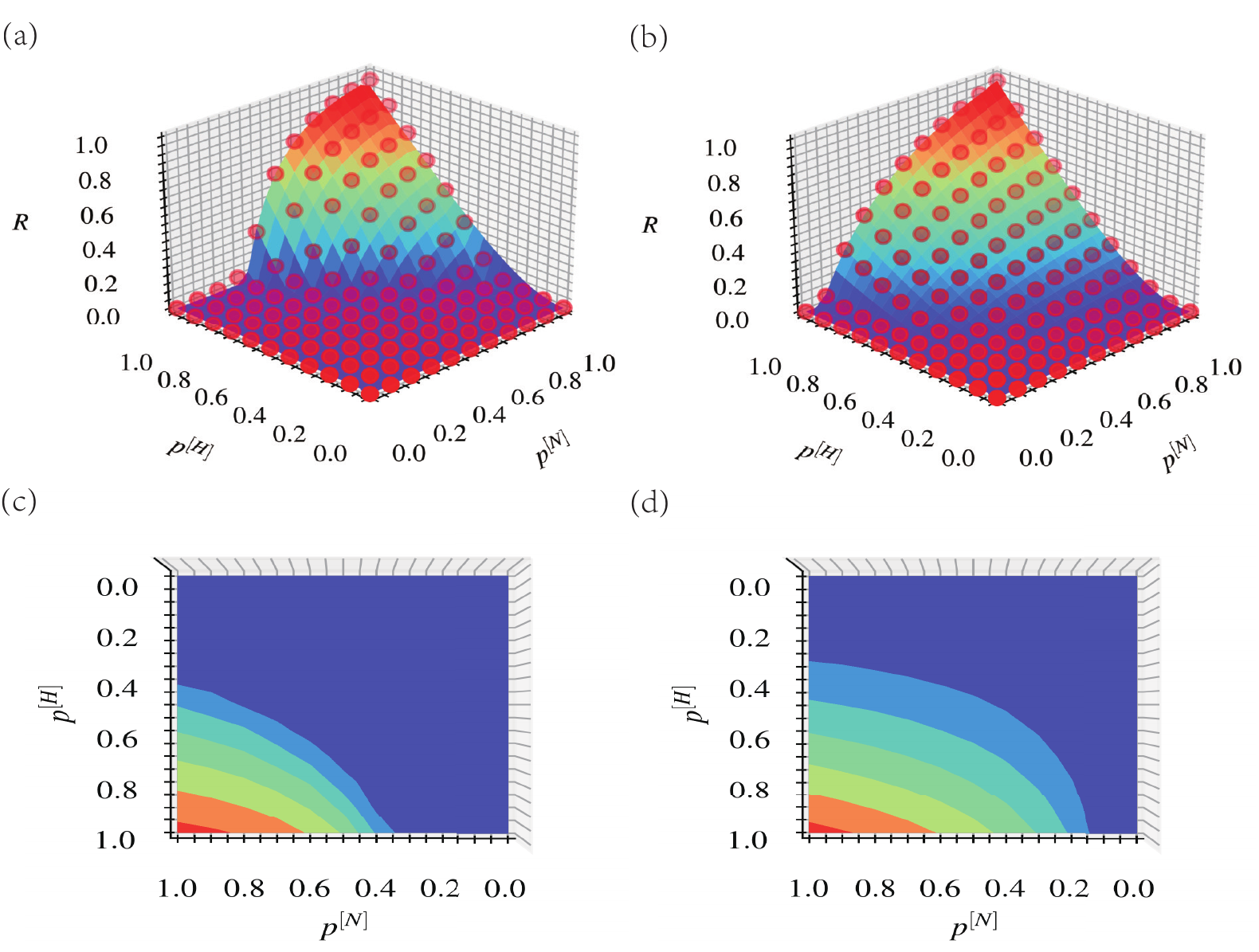}
\caption{Explore the relationship between $p^{[N]}$, $p^{[H]}$ and $R$ . Panel (a) represents a three-dimensional percolation graph on a random hypergraph network with the node size of $N$ = 5,000, the average hyperdegree of 3, the fixed hyperedge cardinality of 2,  and the addition of 2,500 loops. Panel (b) represents the three-dimensional graph of percolation on the scale-free hypergraph network with the node size of $N$ = 5,000, the average hyperdegree of 3, the fixed hyperedge cardinality of 2, $\lambda$ = 2.5, and the addition of 3,750 loops. The independent variables are $p^{[N]}$ and $p^{[H]}$, and the dependent variable is $R$. The surface represents the results under the message passing approach, and the red symbol represents the simulation results. We can see that the two correspond well. To see more clearly how the changes of $p^{[N]}$ and $p^{[H]}$ affect $R$, we show the projection of panels (a) and (b) on the XOY plane through panels (c) and (d) respectively. The results show that $GCC$ is not 0 only when the values of $p^{[N]}$ and $p^{[H]}$ meet certain conditions.}
\label{fig8}
\end{figure*}

where $S_{\mathrm{mp}}$ is the result obtained under the message passing approach. Under different ratios of added loops, Fig.\ref{fig3} (c) and (d) show the influence of the four loops on the experimental accuracy under site percolation and hyperedge percolation, respectively. We can see from the median in the box chart that, in site percolation, 'Type $\uppercase\expandafter{\romannumeral4}$' type of loop has the greatest impact on the accuracy of the network, 'Type $\uppercase\expandafter{\romannumeral1}$' type of loop has the second most impact, and 'Type $\uppercase\expandafter{\romannumeral2}$' type of loop has the least impact. In hyperedge percolation, 'Type $\uppercase\expandafter{\romannumeral1}$' type of loop has the greatest impact on the accuracy of the network, 'Type $\uppercase\expandafter{\romannumeral3}$' type of loop has the second most impact, and 'Type $\uppercase\expandafter{\romannumeral2}$' type of loop has the least impact. When there is the same number of loops in the hypergraph network, we will average the deviation values generated by the four types of loops under the mean-field approach as $\Delta_{mf}$ and average the deviation values caused by the four types of loops under the message passing approach as $\Delta_{mp}$. The results of site percolation and hyperedge percolation with the change of the number of different loops are shown in Fig.\ref{fig3} (e) and (f). It is not difficult to see that our message passing approach is superior to the mean-field method.

Finally, to clarify the relationship between $p^{[N]}$, $p^{[H]}$ and $R$ on the hypergraph network with loops, we have made two three-dimensional graphs that can reflect the relationship between the three as shown in Fig.\ref{fig8} (a) and (b) To see the changes of $p^{[N]}$ and $p^{[H]}$ more clearly, we will show the projection of the surfaces of Fig.\ref{fig8} (a) and (b) on the XOY plane with Fig.\ref{fig8} (c) and (d). It is not difficult to see that when $p^{[N]}$ ($p^{[H]}$) is less than the critical value $p^{[N]}_{c} $($p^{[H]}_ {c} $), regardless of the value of $p^{[H]}$ ($p^{[N]}$), the size of $GCC$ in the network is 0. Only when both $p^{[N]}$ and $p^{[H]}$ are greater than their respective critical values and meet a certain relationship (see Eq.(\ref{10}) for details), can the size of $GCC$ in the network not be zero. That is to say, when the value of $p^{[N]}$ (or $p^{[H]}$) is below its critical value, regardless of the value of $p^{[H]}$ (or $p^{[N]}$), the hypergraph network will be in a state of complete collapse. At this time, the robustness of hypergraph network is very poor. Only when $p^{[N]}$ and $p^{[H]}$ are greater than their critical values, the robustness of the hypergraph network will become more robust with the increase of $p^{[N]}$ or $p^{[H]}$ values.

\subsection{Real-world Network}
The hypergraph network involved previously is our artificially synthesized hypergraph network. In this subsection, we have verified our results in three real scenarios. The construction algorithm of real-world hypergraph networks and more detailed information on these hypergraph networks can be found in Ref.\cite{peng2022targeting}. We used three datasets as follows \cite{SocioPatterns}:
 \begin{figure*}[t]
\centering
\includegraphics[width=1.0\textwidth]{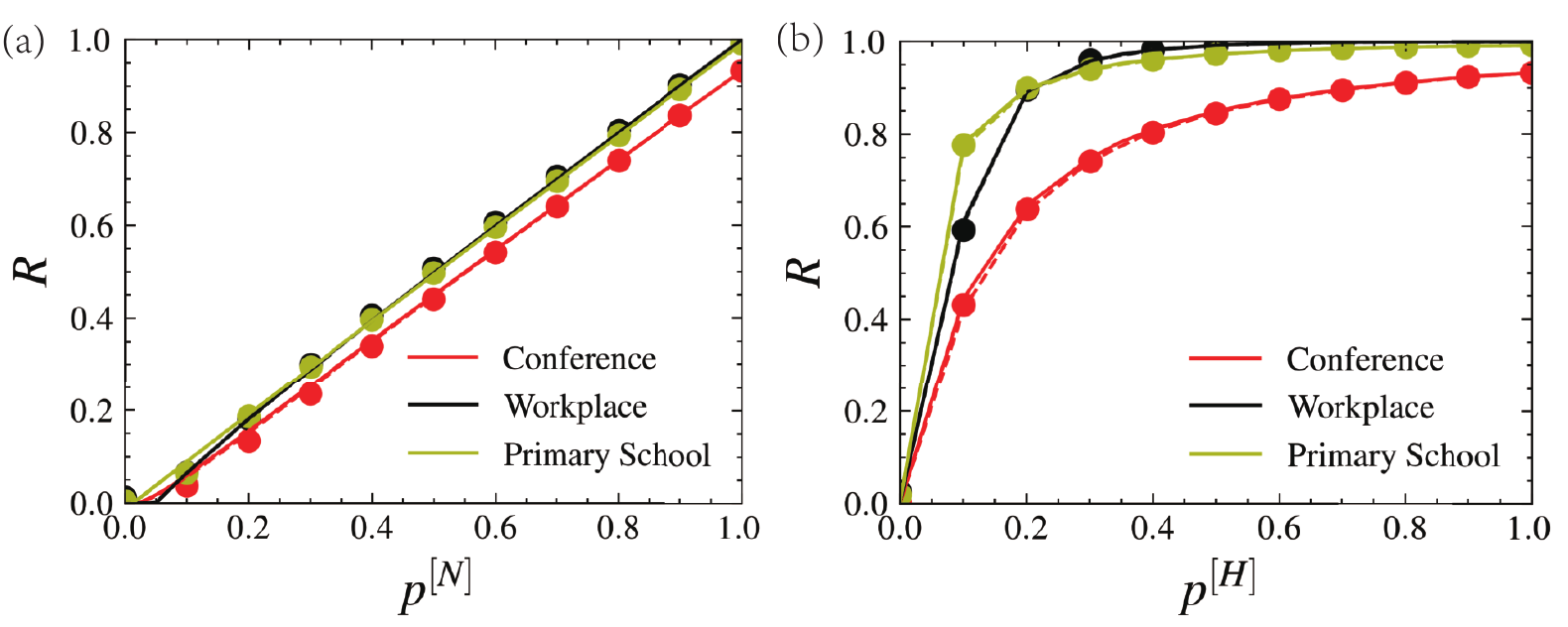}
\caption{Message passing on the real-world hypergraph network. Panels (a) and (b) show the site percolation and hyperedge percolation results on the three real-world hypergraph networks. The solid line represents the result under the mean-field approach, the dotted line represents the effect under the message passing approach, and the symbol represents the simulation result. It can be seen that our proposed approach also performs well on the real-world hypergraph network.}
\label{fig9}
\end{figure*}
\begin{itemize}
\item \textbf{Contacts in the workplace}: This dataset's hypergraph network comprises 92 nodes and 662 hyperedges. Among them, the cardinality of the hyperedge is 2, 3, 4, and 5, the number of which is 618, 42, 2, and 0, respectively \cite{genois2015data}.
\item \textbf{Primary school temporal network data}: This dataset's hypergraph network comprises 242 nodes and 1917 hyperedges. Among them, the cardinality of the hyperedge is 2, 3, 4, and 5, the number of which is 230, 1379, 299, and 9, respectively \cite{gemmetto2014mitigation,stehle2011high}.
\item \textbf{SFHH conference data set}: This dataset's hypergraph network comprises 403 nodes and 1436 hyperedges. Among them, the cardinality of the hyperedge is 2, 3, 4, and 5, the number of which is 494, 739, 175, and 28, respectively \cite{genois2018can}.
\end{itemize}

On the three real-world hypergraph networks mentioned above, we use the mean-field approach, message passing approach and numerical simulation to test these networks, respectively, in which the site percolation results are shown in Fig.\ref{fig9}(a), and the hyperedge percolation results are shown in Fig.\ref{fig9}(b). The experimental results show that our approach not only performs well in the synthetic hypergraph network, but also in the real-world hypergraph network.

	\section{Discussion}
	\label{E}
	In this paper, we investigate random hypergraph models with higher-order interactions. The advantage of this model is that the hyperdegree of nodes and the cardinality of hyperedges can obey any degree distribution, so the model is generalized. We have shown that the same result can be achieved using message passing techniques, compared to using traditional approach (mean-field approach) to compute generating functions using self-consistent equations to solve the scale of the $GCC$ after cascading failures in the loopless network. We provide precise analytical solutions for various properties of the resulting random hypergraph network, including the scale of $GCC$ and the location of percolation phase transitions on random hypergraph networks. We have verified our conclusion on the synthetic random hypergraph network and the real-world hypergraph network.

Secondly, we further find a deviation between the mean-field approach and the true value in the hypergraph network with loops due to the influence of the loops formed by the borders of different hyperedges on the theoretical analysis results. Through experiments, we find that the loop (Type $\uppercase\expandafter{\romannumeral2}$) composed of two general edges and a higher-order edge is closer to the true value. However, whether it is site percolation or hyperedge percolation, the proposed message passing approach performs better than the current mean-field approach under the same conditions. We must also exclude the impact of different network sizes on the experimental results.

In conclusion, we provide an alternative perspective to explore the percolation problem on hypergraph networks. The message passing approach has the advantage that compared to the Monte-Carlo simulation, the experimental results on any specific hypergraph network are not many repeated experiments, which significantly saves the reasonable cost. In addition, the proposal of this scheme also provides a certain idea for the follow-up research on percolation on the directed hypergraph network. Finally, inspired by the percolation problem on simple multilayer networks, how to pass messages on multilayer hypergraph networks is still an exciting problem worthy of study.
	 	
\section*{Data Availability Statement}
The data supporting this study's findings are openly available in sociopatterns at http://www.sociopatterns.org.

\section*{ACKNOWLEDGMENTS}
This work was supported by the National Natural Science Foundation of China under Grant Nos. 62072412, 61902359, 61702148, and 61672468, in part by the Opening Project of Shanghai Key Laboratory of Integrated Administration Technologies for Information Security under grant AGK2018001, and the Natural Science Foundation of Chongqing, No. cstc2021jcyj-msxmX0132.

\section*{REFERENCES}
\bibliography{mybibtex} 

\end{document}